\documentclass[reprint,
 amsmath,amssymb,
 aps,
]{revtex4-1}

\usepackage{graphicx}
\usepackage{dcolumn}
\usepackage{bm}
\usepackage{xcolor}

\begin{document}

\preprint{APS/123-QED}

\title{Accelerating and decelerating space-time optical wave packets in free space}

\author{Murat Yessenov}
\author{Ayman F. Abouraddy}%
\affiliation{CREOL, The College of Optics \& Photonics, University of Central Florida, Orlando, FL 32816, USA}%

\date{\today}

\begin{abstract}
Although a plethora of techniques are now available for controlling the group velocity of an optical wave packet, there are very few options for creating accelerating or decelerating wave packets whose group velocity varies controllably along the propagation axis. Here we show that `space-time' wave packets in which each wavelength is associated with a prescribed spatial bandwidth enable the realization of optical acceleration and deceleration in free space. Endowing the field with precise spatio-temporal structure leads to group-velocity changes as high as $\sim c$ observed over a distance of $\sim20$~mm in free space, which represents a boost of at least $\sim4$ orders of magnitude over X-waves and Airy pulses. The acceleration implemented is in principle independent of the initial group velocity, and we have verified this effect in both the subluminal and superluminal regimes.
\end{abstract}

\maketitle

Acceleration refers to the departure from the condition of uniform motion along a straight line. With regards to \textit{optical} fields, there have been to date only few realizations of acceleration. One example is that of Airy beams traveling in space along curved trajectories \cite{Siviloglou07OL,Efremidis19Optica}, whose acceleration has been recently exploited in inducing synchrotron-like radiation \cite{Henstridge18Science}. The other embodiment is that of axially accelerating wave packets -- pulses whose group velocity $\widetilde{v}$ changes along the propagation axis. However, to date attempts at observing accelerating optical wave packets have yielded only minute changes $\Delta\widetilde{v}/c$ (where $c$ is the speed of light in vacuum). One example is Airy pulses \cite{Abdollahpour10PRL,Kaminer11OE} displaying $\Delta\widetilde{v}\sim10^{-4}c$ over a distance of $\sim\!75$~cm in glass \cite{Chong10NP} (in contrast, Airy pulses in water displayed $\Delta\widetilde{v}/\widetilde{v}\approx0.4$ \cite{Fu2015PRL}). Another example is that of X-waves \cite{Lu92IEEEa,Saari97PRL} after introducing wave-front angular dispersion as proposed by Clerici \textit{et al.} \cite{Clerici08OE}, which has yielded accelerations of $\Delta\widetilde{v}\!\sim\!10^{-3}c$ and decelerations of $\Delta\widetilde{v}\!\sim\!3\times10^{-5}c$ over distances of $\sim\!20$~cm \cite{Lukner09OE}.

Synthesizing axially accelerating optical wave packets is predicated on the ability to control their group velocity. Slow/Fast-light systems can significantly vary the group velocity from $c$ in \textit{resonant} materials or structures \cite{Boyd09Science}, but \textit{not} in free space. In a different context, it is well-known that anomalous group-velocity phenomena occur in the focal volume of focused ultrashort pulses \cite{Porras03PRE}, but we are interested here in wave packets that propagate for extended distances. A recent strategy for controlling $\widetilde{v}$ in \textit{free space} relies on endowing the field with precise spatio-temporal coupling. One approach known as the `flying-focus' \cite{SaintMarie17Optica,Froula18NP,Jolly20OE} exploits longitudinal chromatism to vary $\widetilde{v}$ in the focal volume of a lens, but whose spectrum evolves along the propagation axis. Another approach is that of `space-time' (ST) wave packets \cite{Kondakci16OE,Kondakci17NP} in which each spatial frequency underlying the transverse spatial profile is coupled to a single wavelength \cite{Donnelly93PRSLA,Saari04PRE,Longhi04OE,Parker16OE}, resulting in propagation invariance (diffraction-free and dispersion-free) \cite{Saari97PRL,Turunen10PO,FigueroaBook14} at an arbitrary $\widetilde{v}$ \cite{Salo01JOA,Longhi04OE,Zapata06OL,Wong17ACSP2,Kondakci19NC,Bhaduri19Optica,Bhaduri19unpublished,Schepler20unpublished}. Whereas X-waves allow for group velocities that differ from $c$ by only $\sim0.1\%$ in the paraxial regime \cite{Bowlan09OL,Kuntz09PRA,Bonaretti09OE}, ST wave packets exhibit group velocities in the range from $30c$ to $-4c$ \cite{Kondakci19NC}, \textit{without} the narrowband restrictions typical of slow/fast-light systems \cite{Yessenov19unpub}. Crucially, ST wave packets travel rigidly at a fixed $\widetilde{v}$ in free space, including Airy ST wave packets that travel in a straight line \cite{Kondakci18PRL}. A major rethinking of the spatio-temporal structure of ST wave packets is therefore required to produce axially \textit{accelerating} counterparts.

\begin{figure}[b!]
  \begin{center}
  \includegraphics[width=8.4cm]{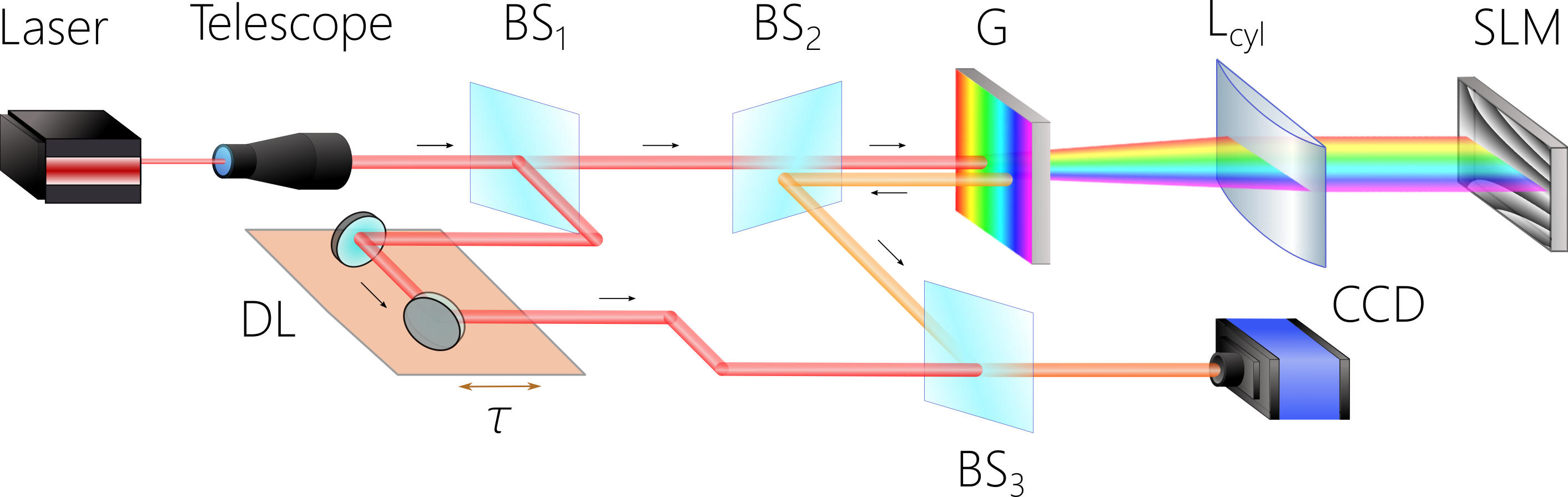}
  \end{center}\vspace{-6mm}
  \caption{Setup for the synthesis and characterization of accelerating ST wave packets. BS: Beam splitter; G: diffraction grating; SLM: spatial light modulator; DL: delay line; L: lens.}
    \label{Fig:ExpSetup}
\end{figure}

\begin{figure*}[t!]
  \begin{center}
  \includegraphics[width=13.4cm]{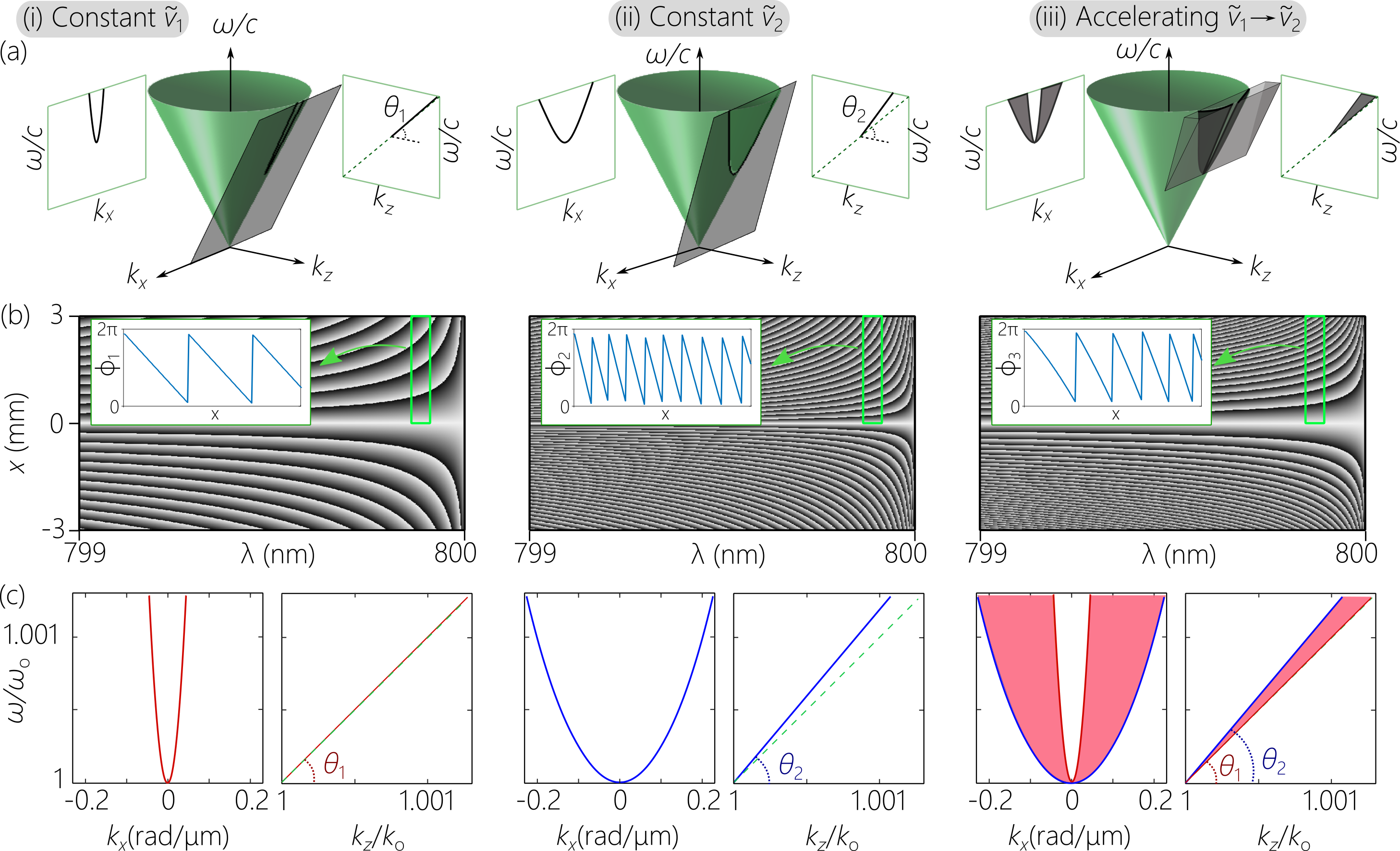}
  \end{center}\vspace{-5mm}
  \caption{(a) Representation of the spectral support domain of ST wave packets on the surface of the light-cone. For propagation-invariant ST wave packets, the support domain is a conic section at the intersection of the light-cone with a tilted spectral plane (here $\theta_{1}\!=\!45.2^{\circ}$ and $\theta_{2}\!=\!47^{\circ}$). For an accelerating ST wave packet, the support domain is the 2D area at the intersection of the light-cone with a wedge-shaped volume bounded by two tilted spectral planes ($\theta_{1}$ and $\theta_{2}$). (b) Phase distribution $\Phi$ imparted by the SLM in Fig.~\ref{Fig:ExpSetup} to the spectrally resolved wave front to produce ST wave packets of group velocity $\widetilde{v}_{1}\!=\!c\tan{\theta_{1}}$, group velocity $\widetilde{v}_{2}\!=\!c\tan{\theta_{2}}$, and acceleration $\widetilde{v}_{1}\rightarrow\widetilde{v}_{2}$. Insets show the phases (modulo $2\pi$) for a fixed temporal frequency (identified by the green box): $\Phi_{1}(x)$ and $\Phi_{2}(x)$ are linear phase distributions each corresponding to a particular spatial frequency; $\Phi_{3}(x)$ is chirped between $\Phi_{1}(x)$ and $\Phi_{2}(x)$, thus corresponding to a finite spatial bandwidth. (c) Spectral projections onto the $(k_{x},\tfrac{\omega}{c})$ and $(k_{z},\tfrac{\omega}{c})$ planes for the ST wave packets in (a) and (b). The projections are 1D curves for constant-$\widetilde{v}$ wave packets, and 2D domains for accelerating wave packets. The dashed green line is the light-line $k_{z}\!=\!\omega/c$.}
  \label{Fig:Concept}
\end{figure*}

Here we show that sculpting the spatio-temporal spectrum of ST wave packets enables the realization of record large axial acceleration. Rather than assigning each spatial frequency to a single wavelength as done previously for propagation-invariant ST wave packets \cite{Donnelly93PRSLA,Yessenov19PRA}, we assign a finite spatial spectrum whose central frequency \textit{and} bandwidth vary in a precisely prescribed manner with the wavelength $\lambda$, which results in an axially encoded $\widetilde{v}$. Using this approach we observe group-velocity changes as large as $\Delta\widetilde{v}\!\sim\!c$ over a distance of $\sim\!20$~mm, representing more than a 4-orders-of-magnitude boost over previous observations. We verify that acceleration and deceleration can each be realized in the subluminal \textit{and} superluminal regimes. Such versatile capabilities in manipulating laser pulses may find applications ranging from nonlinear optics to plasma physics.

To synthesize ST wave packets, we make use of the setup depicted in Fig.~\ref{Fig:ExpSetup}. This arrangement is similar to that used previously is synthesizing wave packets of fixed $\widetilde{v}$, but we extend it here to the synthesis of accelerating and decelerating ST wave packets. A pulsed laser beam from a mode-locked femtosecond Ti:sapphire oscillator is expanded spatially before being directed to a 2D pulse-synthesizer that introduces a judicious spatio-temporal spectral structure into the field via a 2D phase-only spatial light modulator (SLM) \cite{Weiner09Book}. The synthesized ST wave packets travel at arbitrary prescribed $\widetilde{v}$ (without violating relativistic causality \cite{Shaarawi95JMP,SaariPRA18,Saari19PRA,Saari20PRA}), which can be tuned by modifying the phase distribution $\Phi$ imparted to the spatially resolved spectrum by the SLM \cite{Kondakci19NC}. The ST wave packet is then characterized in the spectral domain to ensure the presence of spatio-temporal correlation and in physical space by recording the time-averaged intensity to confirm the diffraction-free propagation. The group velocity is measured via a Mach-Zehnder interferometer with a short plane-wave pulse as the reference (see Supplementary for further details).

\begin{figure*}[t!]
  \begin{center}
  \includegraphics[width=13.8cm]{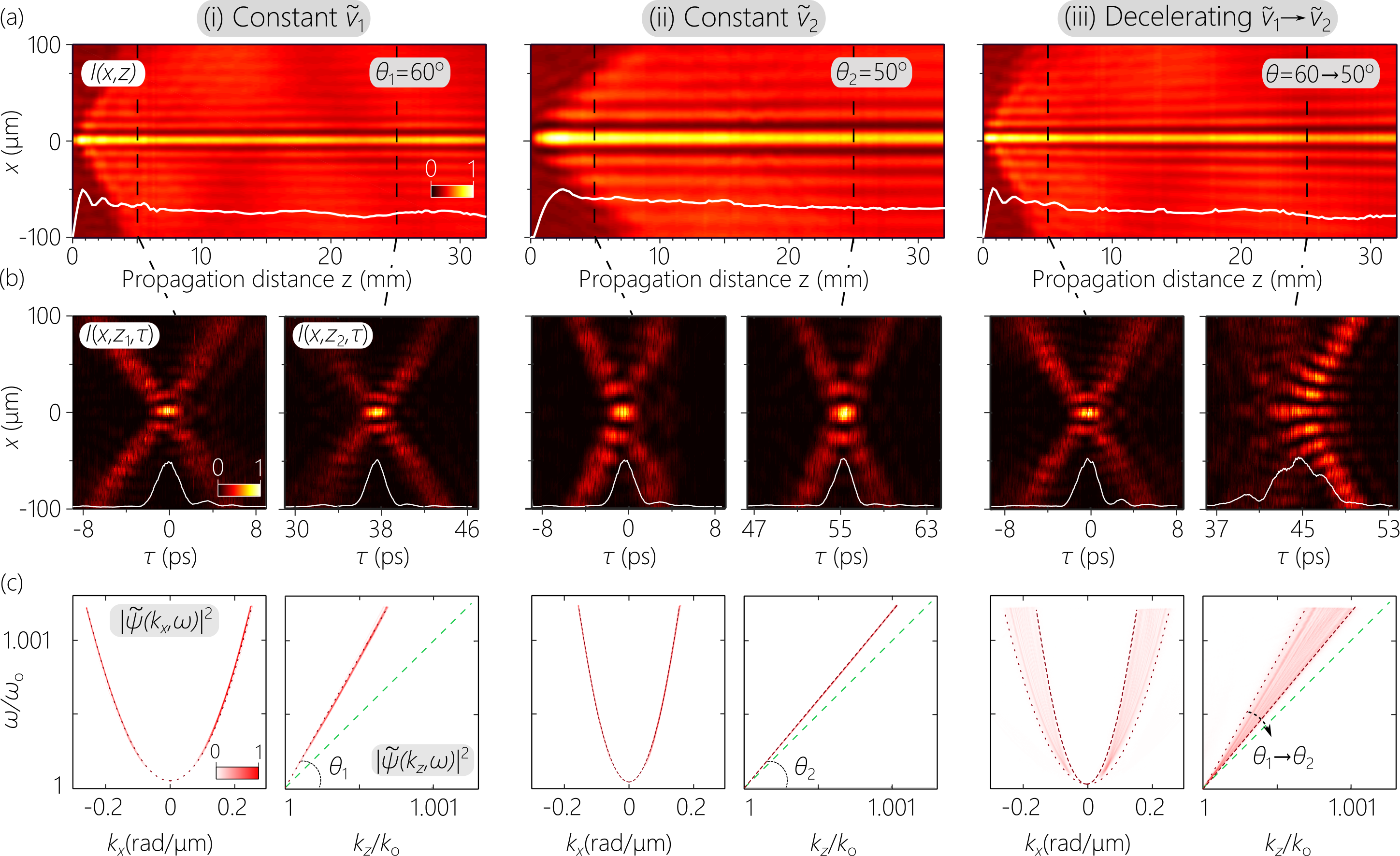}
  \end{center}\vspace{-5mm}
  \caption{(a) Measured axial evolution of the time-averaged intensity $I(x,z)$ for two constant-$\widetilde{v}$ ST wave packets ($\widetilde{v}_{1}\!=\!c\tan{\theta_{1}}$ and $\widetilde{v}_{2}\!=\!c\tan{\theta_{2}}$), and a decelerating ST wave packet $\widetilde{v}_{1}\rightarrow\widetilde{v}_{2}$ ($\widetilde{v}_{2}\!<\!\widetilde{v}_{1}$); $\theta_{1}\!=\!60^{\circ}$ and $\theta_{2}\!=\!50^{\circ}$. The white line represents the normalized on-axis intensity evolution in $z$. (b) Measured spatio-temporal intensity profiles $I(x,z;\tau)$ of the ST wave packets at $z_{1}\!=\!5$~mm and $z_{2}\!=\!25$~mm, corresponding to the marked axial planes in (a). (c) Measured spatio-temporal spectral projections onto the $(k_{x},\omega)$ and $(k_{z},\tfrac{\omega}{c})$ planes. The dotted red curves are theoretical predictions for the constant-$\widetilde{v}$ ST wave packets, the dashed green line is the light-line $k_{z}\!=\!\tfrac{\omega}{c}$, and $\omega_{\mathrm{o}}\!=\!\tfrac{2\pi c}{\lambda_{\mathrm{o}}}$, where $\lambda_{\mathrm{o}}\!\approx\!800$~nm. Theoretical predictions for the decelerating ST wave packet are provided in the Supporting Materials.}
  \label{Fig:AxialMeasurements}
\end{figure*}

The concept of propagation-invariant ST wave packets can be elucidated by considering a single transverse coordinate $x$ (field is uniform along $y$) and an axial coordinate $z$. The free-space dispersion relationship $k_{x}^{2}+k_{z}^{2}\!=\!(\tfrac{\omega}{c})^{2}$ holds; where $k_{x}$ is the transverse component of the wave vector (or `spatial frequency'), $k_{z}$ is the axial component, and $\omega$ is the angular frequency (or `temporal frequency'). This relationship corresponds to the surface of the `light-cone' [Fig.~\ref{Fig:Concept}(a)], and a pulsed optical beam is represented by a 2D spectral support domain on its surface. In contrast, the spectral support for a propagation-\textit{invariant} ST wave packet is a 1D conic section at the intersection of the light-cone with the spectral plane $\Omega\!=\!(k_{z}\!-\!k_{\mathrm{o}})c\tan{\theta}$, whose projection onto the $(k_{z},\tfrac{\omega}{c})$-plane is a straight line making an angle $\theta$ (the spectral tilt angle) with the $k_{z}$-axis [Fig.~\ref{Fig:Concept}(a)-i]; here $\Omega\!=\!\omega-\omega_{\mathrm{o}}$, $\omega_{\mathrm{o}}$ is a fixed frequency, and $k_{\mathrm{o}}\!=\!\tfrac{\omega_{\mathrm{o}}}{c}$ the corresponding wave number. The envelope of the field $E(x,z;t)\!=\!e^{i(k_{\mathrm{o}}z-\omega_{\mathrm{o}}t)}\psi(x,z;t)$ is \cite{Kondakci19NC}:
\begin{equation}
\psi(x,z;t)=\!\!\int\!\!d\Omega\,\widetilde{\psi}(\Omega)e^{i\{k_{x}x-\Omega(t-\tfrac{z}{\widetilde{v}})\}}=\psi(x,0;t\!-\!\tfrac{z}{\widetilde{v}}),
\end{equation}
corresponding to a wave packet traveling rigidly at a group velocity $\widetilde{v}\!=\!\tfrac{\partial\omega}{\partial k_{z}}\!=\!c\tan{\theta}$; where $\widetilde{\psi}(\Omega)$ is the Fourier transform of $\psi(0,0;t)$. Here $k_{x}$ is \textit{not} an independent variable, but is instead related to $\Omega$ and $\theta$ through $k_{x}\!=\!\sqrt{2\omega_{\mathrm{o}}\Omega(1-\widetilde{n})}/c$, where $\widetilde{n}\!=\!\cot{\theta}\!=\!c/\widetilde{v}$ is the group index. Judicious design of the SLM phase distribution enables tuning $\theta$ [Fig.~\ref{Fig:Concept}(b)-i,ii], resulting in ST wave packets with prescribed $\widetilde{v}$ \cite{Kondakci19NC,Bhaduri19Optica,Yessenov19OE}.

Consider two propagation-invariant ST wave packets of group velocities $\widetilde{v}_{1}\!=\!c\tan{\theta_{1}}$ and $\widetilde{v}_{2}\!=\!c\tan{\theta_{2}}$. Their spectral projections onto the $(k_{z},\tfrac{\omega}{c})$-plane are straight line making angles $\theta_{1}$ or $\theta_{2}$ with the $k_{z}$-axis, respectively [Fig.~\ref{Fig:Concept}(c)-i,ii]. We aim to produce a ST wave packet that accelerates from an initial value $\widetilde{v}_{1}$ at $z\!=\!0$ to $\widetilde{v}_{2}$ at $z\!=\!L$: $\widetilde{v}(z)\!=\!\widetilde{v}_{1}+(\widetilde{v}_{2}-\widetilde{v}_{1})z/L\!=\!c/\widetilde{n}(z)$; i.e., the effective group index changes along $z$. To synthesize such a wave packet, we express the relationship between its spatial and temporal frequencies by the ansatz $k_{x}(\Omega,z)\!=\!\sqrt{2\omega_{\mathrm{o}}\Omega(1-\widetilde{n}(z))}/c$. Consequently, each temporal frequency $\Omega$ is associated \textit{not} with a single $k_{x}$ (a condition necessary for propagation invariance), but instead with a \textit{finite} $\Omega$-dependent spatial bandwidth extending over the span $[k_{x}(\Omega,0),k_{x}(\Omega,L)]$. The envelope for the accelerating ST wave packet is given by
\begin{equation}
\psi(x,z;t)=\iint\!dk_{x}d\Omega\widetilde{\psi}(\Omega)\widetilde{h}(k_{x},\Omega)e^{-i\Omega t}e^{i(k_{x}x+k_{z}z)},
\end{equation}
where $\widetilde{h}(k_{x},\Omega)$ imposes a $\Omega$-dependent spatial spectrum. Such a configuration can be realized experimentally by \textit{chirping} the SLM phase distribution associated with each $\Omega$ to correspond to the requisite span of spatial frequencies [Fig.~\ref{Fig:Concept}(b)]. Using this approach, one can realize a ST wave packet whose $\widetilde{v}$ changes between any two terminal values $\widetilde{v}_{1}$ at $z\!=\!0$ and $\widetilde{v}_{2}$ at $z\!=\!L$ (accelerating when $\widetilde{v}_{2}\!>\!\widetilde{v}_{1}$ and decelerating when $\widetilde{v}_{2}\!<\!\widetilde{v}_{1}$). In this case, the spectral projection onto the $(k_{x},\tfrac{\omega}{c})$ plane is a 2D domain [Fig.~\ref{Fig:Concept}(c)-iii] bounded by the 1D curves corresponding to the ST wave packets of the terminal values $\widetilde{v}_{1}$ [Fig.~\ref{Fig:Concept}(c)-i] and $\widetilde{v}_{2}$ [Fig.~\ref{Fig:Concept}(c)-ii]. Similarly, the projection onto the $(k_{z},\tfrac{\omega}{c})$-plane is a 2D wedge-shaped domain bounded by straight lines making angles $\theta_{1}$ and $\theta_{2}$ with the $k_{z}$-axis.

By implementing the appropriate SLM phase distributions experimentally, we obtain the measurement results plotted in Fig.~\ref{Fig:AxialMeasurements}. We set $\theta_{1}\!=\!60^{\circ}$ ($\widetilde{v}_{1}\!\approx\!1.73c$) and $\theta_{2}\!=\!50^{\circ}$ ($\widetilde{v}_{2}\!\approx\!1.19c$), corresponding to a decelerating wave packet. The intensity $I(x,z)\!=\!\int\!dt|\psi(x,z;t)|^{2}$ captured by a CCD camera scanned along $z$ [Fig.~\ref{Fig:ExpSetup}] is provided in Fig.~\ref{Fig:AxialMeasurements}(a) for three ST wave packets: propagation-\textit{invariant} wave packets traveling at $\widetilde{v}_{1}$ and $\widetilde{v}_{2}$, and a \textit{decelerating} wave packet ($\widetilde{v}_{1}\rightarrow\widetilde{v}_{2}$). The axial evolution of the intensity profile shows diffraction-free behavior for all three wave packets, with the axial range for the decelerating wave packet (48~mm) intermediate between the terminal constant-$\widetilde{v}$ wave packets (36~mm for $\theta_{1}\!=\!60^{\circ}$ and 145~mm for $\theta_{2}\!=\!50^{\circ}$ \cite{Yessenov19OE}). Next, time-resolved measurements are carried out along $z$ to assess the group delay $\Delta\tau$ and local group velocity $\widetilde{v}(z)$. We plot in Fig.~\ref{Fig:AxialMeasurements}(b) the spatio-temporal profile $I(x,z;\tau)\!=\!|\psi(x,z;\tau)|^{2}$ at two axial positions ($z\!=\!5$~mm and 25~mm) for each wave packet. Over this distance $\Delta\tau\!\sim\!38$~ps and $\Delta\tau\!\sim\!55$~ps for the ST wave packets traveling at $\widetilde{v}_{1}$ and $\widetilde{v}_{2}$, respectively, which are consistent with $\tfrac{\Delta z}{v_{1}}$ and $\tfrac{\Delta z}{v_{2}}$ ($\Delta z\!=\!20$~mm); while $\Delta\tau$ for the wave packet decelerating between the values has an intermediate value. The measured spectral projections onto the $(k_{x},\tfrac{\omega}{c})$ and $(k_{z},\tfrac{\omega}{c})$ planes confirm that the synthesized ST wave packets reproduce the target spectral structure. Projections onto the $(k_{x},\tfrac{\omega}{c})$ plane are 1D parabolic curves for the constant-$\widetilde{v}$ wave packets and is a 2D domain for the decelerating wave packet bounded by the 1D curves for the constant-$\widetilde{v}$ limits. Projections onto the $(k_{z},\tfrac{\omega}{c})$-plane are straight lines for the constant-$\widetilde{v}$ wave packets (making angles of $60^{\circ}$ and $50^{\circ}$ with the $k_{z}$-axis), and a wedge-shaped domain extending between $60^{\circ}$ and $50^{\circ}$ for the decelerating wave packet.

For a constant-$\widetilde{v}$ ST wave packet, the propagation distance is $L_{\mathrm{max}}\!=\!\tfrac{c}{\delta\omega}\tfrac{1}{|1-\cot{\theta}|}$, where the spectral uncertainty $\delta\omega$ (which is typically much smaller than the full temporal spectral bandwidth) is the unavoidable `fuzziness' in the association between the each $k_{x}$ and $\omega$ (due in our setup mainly to the finite aperture of the diffraction grating that limits its spectral resolution) \cite{Yessenov19OE}. In the case of accelerating ST wave packets, because we no longer have a one-to-one relationship between $k_{x}$ and $\omega$, pulse dispersion is introduced \cite{Kondakci19OL}; see Fig.~\ref{Fig:AxialMeasurements}(b)-iii.

\begin{figure}[t!]
  \begin{center}
  \includegraphics[width=8.6cm]{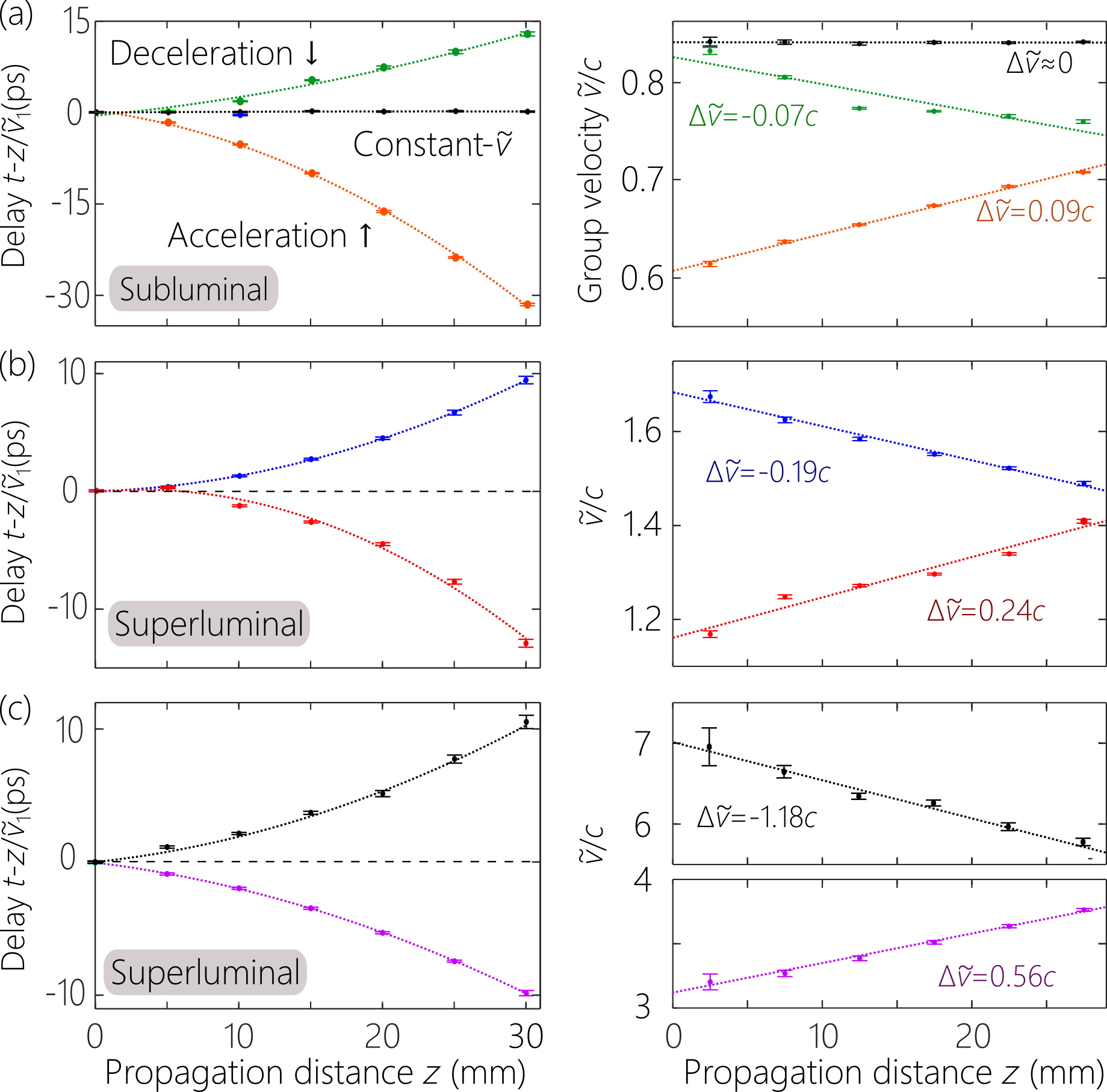}
  \end{center}\vspace{-5mm}
  \caption{(a) Measured group delays (left) with respect to a pulse traveling at the initial group velocity $\widetilde{v}_{1}$ at $z\!=\!0$, and group velocities (right). We plot data for accelerating and decelerating \textit{subluminal} ST wave packets over a propagation distance of 30~mm, and also plot for comparison data for a constant-$\widetilde{v}$ ST wave packet at the initial group velocity $\widetilde{v}_{1}$. (b) Same as (a) for superluminal ST wave packets. (c) Same as (b) for superluminal ST wave packets with large acceleration and deceleration. Points are the measured data, and the dotted curves are theoretical fits (quadratic for the delays, linear for the group velocities).}\vspace{-2mm}
    \label{Fig:group_velocity}
\end{figure}

To establish that the synthesized ST wave packets accelerate and decelerate at the prescribed rates, we plot in Fig.~\ref{Fig:group_velocity} the measured $\Delta\tau$ (with respect to a frame moving at the initial speed $\widetilde{v}_{1}$) obtained at 5-mm axial intervals and the estimated $\widetilde{v}(z)$ for accelerating and decelerating ST wave packets. Therefore $\Delta\tau$ at $z\!=\!0$ is zero in all cases (and remains zero for constant-$\widetilde{v}$ wave packets), and subsequently $\Delta\tau\!>\!0$ for decelerating wave packets and $\Delta\tau\!<\!$ for accelerating ones. Moreover, the data in Fig.~\ref{Fig:group_velocity} serves also to confirm that acceleration can be realized in both the subluminal ($\theta\!<\!45^{\circ}$) and superluminal ($\theta\!>\!45^{\circ}$) regimes \cite{Yessenov19PRA}. In all cases, we find that $\Delta\tau$ has a \textit{parabolic} dependence and $\widetilde{v}(z)$ a \textit{linear} dependence on $z$, as expected for linearly accelerating wave packets.

We first present in Fig.~\ref{Fig:group_velocity}(a) measurements for \textit{subluminal} decelerating $\Delta\widetilde{v}\!=\!-0.07c$ ($\theta_{1}\!=\!40^{\circ}\rightarrow\theta_{2}\!=\!30^{\circ}$, $\widetilde{v}_{1}\!=\!0.84c\rightarrow\widetilde{v}_{2}\!=\!0.58c$) and accelerating $\Delta\widetilde{v}\!=\!0.09c$ ($30^{\circ}\rightarrow43^{\circ}$, $0.58c\rightarrow0.93c$) wave packets. Next we present in Fig.~\ref{Fig:group_velocity}(b) measurements for \textit{superluminal} decelerating $\Delta\widetilde{v}\!=\!-0.19c$ ($70^{\circ}\rightarrow50^{\circ}$, $2.75c\rightarrow1.19c$) and accelerating $\Delta\widetilde{v}\!=\!0.24c$ ($50^{\circ}\rightarrow70^{\circ}$, $1.19c\rightarrow2.75c$) wave packets. Finally, Fig.~\ref{Fig:group_velocity}(c) provides further measurements for \textit{superluminal} decelerating $\Delta\widetilde{v}\!=\!-1.18c$ ($82^{\circ}\rightarrow70^{\circ}$, $7.12c\rightarrow2.74c$) and accelerating $\Delta\widetilde{v}\!=\!0.56c$ ($70^{\circ}\rightarrow82^{\circ}$, $2.74c\rightarrow7.12c$) wave packets, which exceeds by 4-orders-of-magnitude previously reported measurements \cite{Lukner09OE,Chong10NP}.

The synthesized accelerating ST wave packets fall short of the targeted terminal group velocities $\widetilde{v}_{2}$ due to the finite pixel size and number of pixels of the SLM that limit the achievable chirping rate. This is associated with pulse distortion in the accelerating ST wave packets (compared to the rigid propagation of the constant-$\widetilde{v}$ wave packets); see Fig.~\ref{Fig:AxialMeasurements}(c) confirmed by the simulations provided in the Supporting Materials. This limitation can be alleviated by utilizing lithographically inscribed phase plates of higher spatial resolution and larger size \cite{Wang15ProgPh,Kondakci18OE,Bhaduri19OL,Yessenov19Optica}. 

The strategy outlined here for the synthesis of accelerating and decelerating optical wave packets represents a fundamental departure from previous approaches, such as Airy pulses or wavefront-modulation X-waves. Our approach is quite general and does not need a resonant material or structure, and thus has no fundamental limit on the exploitable bandwidth; it does not require utilizing a dispersive medium; and it can be utilized to produce other axial acceleration profiles such as sub-linear or super-linear acceleration, or combined acceleration \textit{and} deceleration over prescribed axial domains.

It has been shown that accelerating fields are can help selectively manipulate nonlinear frequency conversion processes when combined with quasi-phase-matching schemes \cite{Bahabad11PRA}. The results here provide such acceleration in a versatile experimental arrangement, and may also lead to new sources of radiation \cite{Henstridge18Science}, acceleration of charged particles \cite{Jolly19OL}, in addition to enhancements in nonlinear optics \cite{Bahabad11PRA,Hu17SR,Jia19PRL} and plasma interactions \cite{Howard19PRL}. By synthesizing multiple co-propagating ST wave packets in the same or different spectral windows \cite{Yessenov19unpub}, co-propagating pulses can be arranged to collide once or multiple times over the course their propagation. Recent work indicates that ST wave packets can be coupled into planar waveguides \cite{Shiri20unpub}, but more research is needed to ascertain whether they can be coupled into optical fibers.

In conclusion we have synthesized a new class of ST wave packets that (1) accelerate or decelerate in free space; (2) can be realized in both the subluminal and superluminal regimes; (3) exhibit record changes in the group velocity extending to $\Delta\widetilde{v}\!\sim\!1.2c$ over a propagation distance of $\sim\!20$~mm; (4) all by virtue of their spatio-temporal spectral structure in which each wavelength is associated with a prescribed spatial spectrum whose center and bandwidth are wavelength-dependent. The achieved accelerations are at least 4~orders-of-magnitude larger than previously reported.

\noindent
\textit{Ackowledgments}.--- This work was funded by the U.S. Office of Naval Research contract N00014-17-1-2458.

\newpage

\bibliography{Main}

\end{document}


\preprint{APS/123-QED}

\title{Supplemental Material for ``Accelerating and decelerating space-time optical wave packets in free space''}

\author{Murat Yessenov}
\author{Ayman F. Abouraddy}%
\affiliation{CREOL, The College of Optics \& Photonics, University of Central Florida, Orlando, FL 32816, USA}%

\date{\today}

\renewcommand{\thepage}{S\arabic{page}}  
\renewcommand{\thesection}{S\arabic{section}}   
\renewcommand{\thefigure}{S\arabic{figure}}

\maketitle
\section{Experimental setup}

To make our paper fully self-contained, we have provided here the experimental details of the synthesis and characterization of ST wave packets utilized in observing axial acceleration and deceleration. This includes the details of the experimental arrangement, the design of the phase distributions required to synthesize accelerating and decelerating ST wave packets, and the characterization setups.

\subsection{ST wave packets synthesis}

Space-time (ST) wave packets are synthesized using the 2D pulse-shaper depicted in Fig.~1 of the main text. Starting with femtosecond laser pulses at a wavelength of $\sim800$~nm from a Ti:sapphire laser (Tsunami, Spectra Physics), we expand the transverse beam size to a diameter of 25~mm (defined as full-width at half maximum, FWHM) via a telescope system (Thorlabs BE02-05-B, optical beam expander). The spectrum of this pulsed beam is resolved in space by a diffraction grating (having a ruling density of 1200~lines/mm and of area $25\times25$~mm$^2$; Newport 10HG1200-800-1), and then modulated by a reflective phase-only spatial light modulator (SLM; Hamamatsu X10468-02) placed at the focal plane of a collimating cylindrical lens (focal length $f\!=\!50$~cm); see Fig.~1 in the main text.

In contrast to traditional ultrafast pulse shapers that implement a 1D spectral phase modulation \cite{Weiner09Book}, we implement a 2D phase distribution that assigns to each wavelength $\lambda$ a single spatial frequency $k_x$ \cite{Yessenov19OPN}. The retro-reflected wave front is reconstituted by the grating to yield the ST wave packet whose group velocity $\widetilde{v}$ can be tuned by modifying the SLM phase distribution \cite{Kondakci19NC} (without violating relativistic causality \cite{Shaarawi95JMP,SaariPRA18,Saari19PRA,Saari20PRA}). 

\subsection{Designing the SLM phase pattern for fixed group velocity}

The SLM imparts a 2D phase distribution $\Phi(x,y)$ to the spectrally resolved incident wave front, as shown in three examples provided in Fig.~2(b) of the main text. The wavelengths are spread in space along the $y$-axis (the horizontal direction in our experimental configuration), so that each position $y$ is associated with one wavelength $\lambda$. We thus express the phase pattern as $\Phi(x,\lambda)$; the proportionality of $y$ and $\lambda$ is calibrated experimentally. The SLM phase pattern is designed to target a prescribed spatio-temporal spectral correlation between the spatial frequencies $k_{x}$ and wavelengths $\lambda$. When we desire each $k_{x}$ to be associated with a single $\lambda$, as is the case when a ST wave packet of fixed group velocity is desired, then the implemented phase distribution is linear along the vertical direction $x$,
\begin{equation}
\Phi(x,\lambda)=k_{x}(\lambda)x.
\end{equation}
This approach enables us to assign to each wavelength $\lambda$ a spatial frequency $k_x (\lambda)$, according to the narrowband parabolic approximation:
\begin{equation}
\frac{\omega-\omega_{\mathrm{o}}}{\omega_{\mathrm{o}}}=\frac{k_{x}^{2}}{2k_{\mathrm{o}}^{2}}\frac{1}{1-\widetilde{n}};
\end{equation}
here $\omega\!=\!\tfrac{2\pi c}{\lambda}$, $c$ is the speed of light in vacuum, and $\omega_{\mathrm{o}}$ is a fixed frequency (corresponding here to $\lambda_{\mathrm{o}}\!\approx\!800$~nm), and $k_{\mathrm{o}}$ is the corresponding wave number. The group index $\widetilde{n}$ then determines the wave packet group velocity $\widetilde{v}\!=\!c/\widetilde{n}$. We define a spectral tilt angle $\theta$, where $\widetilde{n}\!=\!\cot{\theta}$ and $\widetilde{v}\!=\!c\tan{\theta}$.

\subsection{Designing the SLM phase pattern for accelerating ST wave packets}

In case of accelerating or decelerating ST wave packets, each wavelength $\lambda$ is \textit{no longer} associated with a single spatial frequency $k_{x}$. Instead, each $\lambda$ is to be associated with a \textit{finite} spatial bandwidth, with both the central spatial frequency and the spatial bandwidth change with wavelength in a prescribed manner. We define this wavelength-dependent spatial spectrum in terms of the two limit values of $k_{x}$, $k_{x1}(\lambda)$ and $k_{x2}(\lambda)$. These limits are designed to achieve an acceleration between a group velocity $\widetilde{v}_{1}\!=\!c\tan{\theta_{1}}\!=\!c/\widetilde{n}_{1}$ at $z\!=\!0$ and $\widetilde{v}_{2}\!=\!c\tan{\theta_{2}}\!=\!c/\widetilde{n}_{2}$ at $z\!=\!L$. The limit spatial frequencies $k_{x1}(\lambda)$ and $k_{x2}(\lambda)$ associated with the wavelength $\lambda$ are then given by
\begin{eqnarray}
k_{x1}(\lambda)&=&\sqrt{2\omega_{\mathrm{o}}\Omega(1-\widetilde{n}_{1})}/c,\nonumber\\
k_{x2}(\lambda)&=&\sqrt{2\omega_{\mathrm{o}}\Omega(1-\widetilde{n}_{2})}/c,
\end{eqnarray}
where $\Omega\!=\!\omega-\omega_{\mathrm{o}}$. Consequently, the phase distribution displayed by the SLM imparts a quadratic phase to the incident wave front
\begin{equation}
\Phi(x,\lambda)=k_{x1}(\lambda)+\left[k_{x2}(\lambda)-k_{x1}(\lambda)\right]\frac{x^{2}}{2x_{\mathrm{max}}}
\end{equation}
to each $\lambda$; here $x_{\mathrm{max}}$ is the maximum spatial width of the SLM active area exploited. This corresponds to an axially encoded spatial frequency distribution of the form
\begin{equation}
k_{x}(\lambda,z)=\sqrt{2\omega_{\mathrm{o}}\Omega(1-\widetilde{n}(z))}/c,  
\end{equation}
where the axially varying group index $\widetilde{n}(z)$ is designed to yield a ST wave packet with a linearly changing group velocity $\widetilde{v}(z)$, $\widetilde{v}_{1}\rightarrow\widetilde{v}_{2}$, at a prescribed rate. Initially, the spatio-temporal spectrum corresponds at $z\!=\!0$ to a wave packet with group velocity $\widetilde{v}_{1}$ whose spectral projection onto the $(k_{z},\tfrac{\omega}{c})$-plane is a line making an angle $\theta_{1}$ with the $k_{z}$-axis, and evolves into a wave packet with group velocity $\widetilde{v}_{2}$ whose spectral projection onto the $(k_{z},\tfrac{\omega}{c})$-plane is a line making an angle $\theta_{2}$ with the $k_{z}$-axis. The spectral projection of the accelerating ST wave packet onto the $(k_{z},\tfrac{\omega}{c})$-plane is a wedge-shaped region bounded by the angles $\theta_{1}$ and $\theta_{2}$ with the $k_{z}$-axis. 

The active area of the SLM chip is split along $x$ into two halves: in one half we impress positive spatial frequencies $k_{x}$ while in the other half the corresponding negative frequencies $-k_{x}$. The phase-modulated wave front retroreflects back from the SLM to the diffraction grating, which reconstitutes the pulse and thus produces the ST wave packets with pre-designed acceleration or deceleration. As such, the phase distribution in Fig.~2(b)-iii of the main text is produced. The temporal bandwidth exploited in our experiments is $\Delta\lambda\!\approx\!1$~nm, which corresponds to a pulse width of $\approx\!2.2$~ps (observed at $x\!=\!0$ in the case of ST wave packets).

\subsection{Characterization of ST wave packets}

To confirm the that the desired spatio-temporal spectral structure has been introduced into the optical field, the spectral projection onto the $(k_{x},\lambda)$-plane is measured by a combination of spatial and temporal Fourier transforms. A portion of the field retroreflected from SLM is directed to a spherical lens (focal length $f\!=\!7.5$~cm) after which the intensity is recorded with a CCD camera (The ImagingSource, DMK 33UX178). This arrangement is not shown in Fig.~1 of the main text for simplicity. From this measurement, and along with the free-space dispersion relationship $k_{x}^{2}+k_{z}^{2}\!=\!(\tfrac{\omega}{c})^{2}$, we obtain the spectral projection onto the $(k_{z},\lambda)$-plane. Utilizing this approach, we obtained the projected spectra shown in Fig.~3(c) of the main text.

The axial evolution of the time-averaged intensity distribution of the ST wave packet upon free propagation $I(x,z)\!=\!\int\!dt|\psi(x,z;t)|^{2}$is captured by a CCD camera (The ImagingSource, DMK 27BUP031) scanned along the $z$-axis. This measurement confirms the diffraction-free propagation of the wave packet, as shown in Fig.~3(a) of the main text.

Finally, the spatio-temporal profile of the ST wave packet is measured using two-arm Mach-Zehnder interferometer in which the ST wave packets are synthesized in one arm, and a short reference pulse from the initial Ti:sapphire laser (pulsewidth $\sim\!100$~fs) travels along the other arm that contains an optical delay line $\tau$. When the two wave packets (the ST wave packet and the reference pulse) overlap in space and time, spatially resolved interference fringes are observed by a CCD camera at a fixed axial position $z$ (The ImagingSource, DMK 27BUP031), from whose visibility we extract the spatio-temporal profile at that plane. By scanning the delay $\tau$, we can reconstruct the spatio-temporal intensity profile $I(x,z;\tau)$ at a fixed axial plane $z$. Using this approach, we obtained the wave packet profiles shown in Fig.~3(b) of the main text.

To obtain the local group velocity $\widetilde{v}(z)$ of the group velocity, we measure the group delay incurred by the ST wave packet at different axial positions. We first arrange for the ST wave packet and the reference pulse to overlap at $z\!=\!0$ as described above. We then axially displace the CCD camera to a different axial position $\Delta z$, which results in a loss of interference due to the group-velocity mismatch between the ST wave packet traveling at $\widetilde{v}$ and the reference pulse traveling at $c$. The interference is regained, however, by adjusting the delay by $\Delta\tau$ in the reference arm, from which we obtain the group delay and thence the average group velocity over the distance $\Delta z$, $\widetilde{v}(z)\!=\!\tfrac{\Delta z}{\Delta\tau}$ at each axial location $z$. By determining the center of this fitted wave packet, we can then determine the group delay $\Delta\tau$. At each position $z$, we fit a Gaussian distribution to the temporal intensity profile at the spatial center of the wave packet $I(0,z;\tau)$. Using this strategy, we obtain the data plotted in Fig.~4 of the main text.

\section{Simulation of accelerating and decelerating space-time wave packets}

\begin{figure*}[t!]
  \begin{center}
  \includegraphics[width=17.0cm]{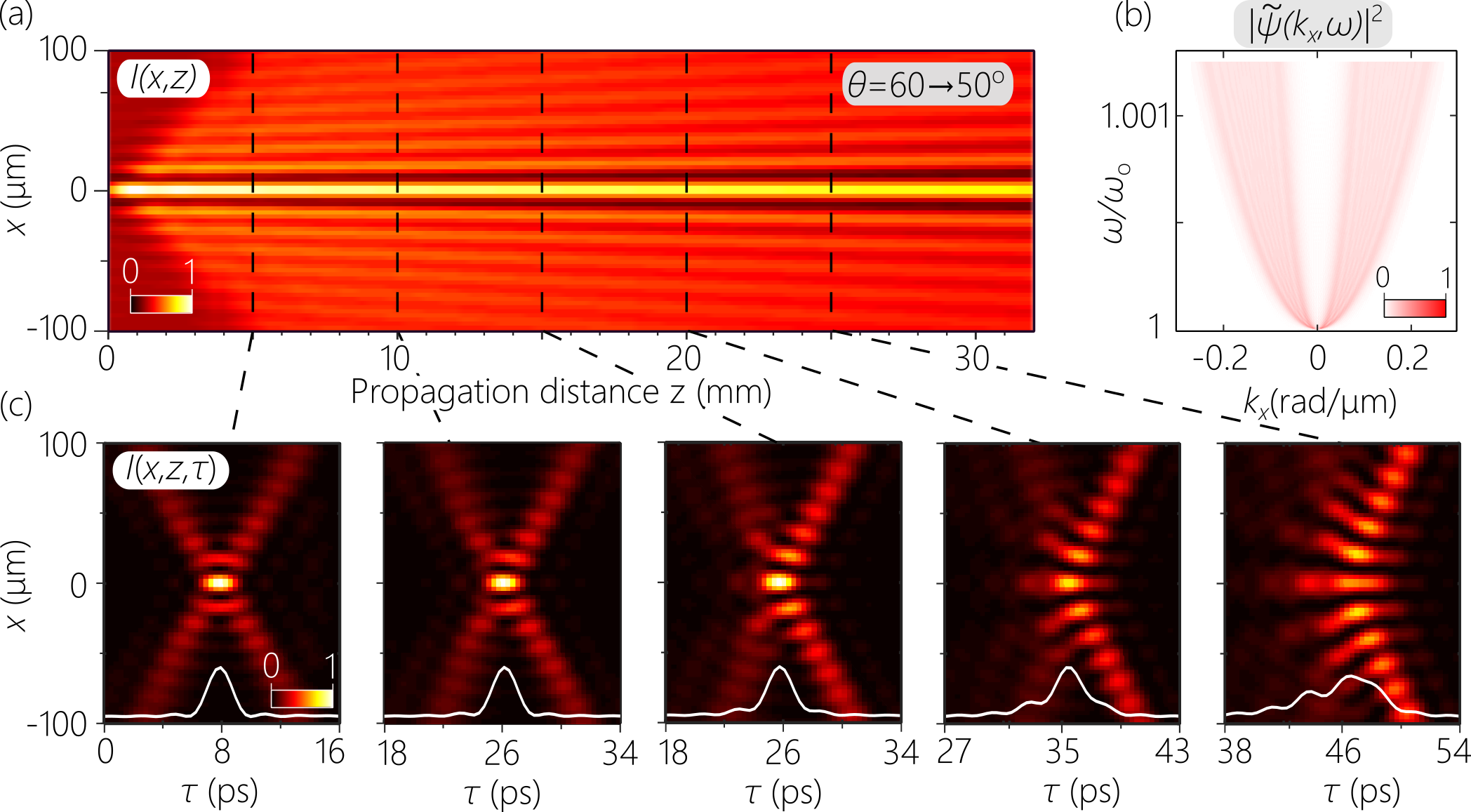}
  \end{center}\vspace{-5mm}
  \caption{Simulations of a decelerating ST wave packet corresponding to a change in the spectral tilt angle of $\theta_{1}\!=\!60^{\circ}\rightarrow\theta_{2}\!=\!50^{\circ}$, corresponding to the experimental results plotted in Fig.~3 of the main text. (a) Axial evolution of the time-averaged intensity $I(x,z)$. (b) Spatio-temporal spectral projection onto the $(k_{x},\tfrac{\omega}{c})$-plane. (c) Spatio-temporal intensity profiles $I(x,z;\tau)$ of the ST wave packet at five axial positions along $z$, from 5~mm to 25~mm, corresponding to the marked axial planes in (a). White line curves in (c) represent on-axis temporal profiles $I(x=0,z;\tau)$ of the ST wave packet at each axial plane.}
  \label{Fig:Simulation}
\end{figure*}

The simulation of ST wave packets is based on the Fourier-transform split-step method -- a well-known technique for pulsed-beam propagation in the Fourier domain \cite{Wartak2013Book}. We start off the simulation with an optical field having a spatio-temporal profile modeling that utilized in the experiment: quasi-plane-wave femtosecond laser pulses having a central wavelength of $\lambda\!\sim\!800$~nm, a temporal bandwidth of $\Delta\lambda\!\sim\!10$~nm, and a transverse spatial width of $\Delta x\!\sim\!20$~mm. In the experiment, the initial laser beam from the Ti:sapphire laser is expanded to this size via a telescope. The spatio-temporal spectrum is initially separable with respect to the spatial and temporal degrees of freedom. We then transform the spatio-temporal spectrum by modulating it with the 2D phase distribution that is imparted in the experiment by the SLM to the spectrally resolved wave front (Fig.~1 of the main text). We also include the effect of the finite aperture size of the SLM active area. This leads to a truncation in the temporal spectrum (reducing the temporal bandwidth from $\Delta\lambda\sim10$~nm to $\approx1$~nm) and a truncation in the transverse spatial extent of the field ($\Delta x\!=\!12$~mm). We then propagate the field in the spatio-temporal Fourier domain $(k_{x},\omega)$ over small axial step ($300~\mu$m) over the whole range of axial propagation along $z$ (30~mm). Finally, we implement a 2D inverse-Fourier transform in space and time to obtain the field profile $\psi(x,z;t)$ along $z$.

We plot in Fig.~\ref{Fig:Simulation} simulation results for a decelerating ST wave packet having parameters corresponding to the experimentally realized ST wave packet in Fig.~3 of the main text: a spectral tilt angle of $\theta_{1}=60^{\circ}\rightarrow\theta_{2}=50^{\circ}$ corresponding to a change in group velocity $\widetilde{v}_{1}=1.73c\rightarrow\widetilde{v}_{2}=1.19c$. The time-averaged intensity $I(x,z)$ is provided in Fig.~\ref{Fig:Simulation}(a), which reveals diffraction-free behaviour over the extended propagation length, in excellent agreement with the experiment observations in Fig.~3(a) in the main text. 

Additionally, we plot the projection onto the $(k_{x},\tfrac{\omega}{c})$-plane of the spatio-temporal spectrum utilized in the simulations [plotted in Fig.~\ref{Fig:Simulation}(b)], which is also in excellent agreement with the measured spectral projections [Fig.~3(c) in the main text]. Finally, in Fig.~\ref{Fig:Simulation}(c) we plot spatio-temporal intensity profiles of the decelerating ST wave packet at selected axial planes along the propagation axis $z$ to examine closer its axial evolution. The intensity profiles $I(x,z;\tau)\!=\!|\psi(x,z;\tau)|^{2}$ at 5 locations along $z$ (from 5~mm to 25~mm in 5-mm axial increments) reveal the progression of the distortion of the wave packet profile, which shows excellent correspondence with the measurements plotted in Fig.~3(b) of the main text. The distortion becomes noticeable only at the end of the propagation distance.

Two factors contribute to the wave packet distortion along the propagation axis. First, strict propagation invariance requires tight spatio-temporal spectral correlations: each spatial frequency must be assigned to a single wavelength or temporal frequency. On the other hand, axial acceleration breaks the propagation invariance, and each wavelength has associated with it a finite wavelength-dependent spatial bandwidth. Although this helps realize the wave packet acceleration, the wave packet will inevitably distort with propagation. Second, in our experiments, the finite size of the SLM pixels (and thus the finite number of pixels within the active area of the SLM) sets a limit on the achievable precision of the phase chirping required to produce the target acceleration. 

\newpage

\bibliography{sup}